\documentclass[iop,tighten,apjl]{emulateapj}
\usepackage[]{times, graphicx}
\newcommand{\pref}{\protect\ref}
\newcommand{\soho}{{\em SOHO{}}}
\newcommand{\sdo}{{\em SDO{}}}

\begin{document}

\shorttitle{Marking the Sun's Giant Convective Scale}

\shortauthors{S.W. McIntosh et al.}

\title{Identifying Potential Markers of the Sun's Giant Convective Scale}

\author{Scott W. McIntosh\altaffilmark{1}, Xin Wang\altaffilmark{1,2}, Robert J. Leamon\altaffilmark{3} \& Philip H. Scherrer\altaffilmark{4}}
\altaffiltext{1}{High Altitude Observatory, National Center for Atmospheric Research, P.O. Box 3000, Boulder, CO 80307}
\altaffiltext{2}{School of Earth and Space Sciences, Peking University, Beijing, 100871, China}
\altaffiltext{3}{Department of Physics, Montana State University, Bozeman, MT 59717}
\altaffiltext{4}{W. W. Hansen Experimental Physics Lab, Stanford University, Stanford, CA 94305}

\begin{abstract}
Line-of-sight magnetograms from the Helioseismic and Magnetic Imager (HMI) of the {\em Solar Dynamics Observatory} (SDO) are analyzed using a diagnostic known as the ``Magnetic Range of Influence,'' or MRoI. The MRoI is a measure of the length over which a photospheric magnetogram is balanced and so its application gives the user a sense of the connective length scales in the outer solar atmosphere. The MRoI maps and histograms inferred from the \sdo/HMI magnetograms primarily exhibit four scales: a scale of a few megameters that can be associated with granulation, a scale of a few tens of megameters that can be associated with super-granulation, a scale of many hundreds to thousands of megameters that can be associated with coronal holes and active regions, and a hitherto unnoticed scale that ranges from 100 to 250 megameters. We infer that this final scale is an imprint of the (rotationally-driven) giant convective scale on photospheric magnetism. This scale appears in MRoI maps as well-defined, spatially distributed, concentrations that we have dubbed ``g-nodes.'' Furthermore, using coronal observations from the Atmospheric Imaging Assembly (AIA) on \sdo{}, we see that the vicinity of these g-nodes appears to be a preferred location for the formation of extreme ultraviolet (EUV, and likely X-Ray) brightpoints. These observations and straightforward diagnostics offer the potential of a near-real-time mapping of the Sun's largest convective scale, a scale that possibly reaches to the very bottom of the convective zone.
\end{abstract}

\section{Introduction}
The region of the solar atmosphere that is hidden from direct observation by its profound optical depth \-- the Sun's convective interior \-- is an ocean of boiling, bubbling plasma sustained from beneath by a rotating, nuclear furnace. The convective layer that forms the last third of the solar interior masks the process, or processes, which govern the production and perpetual eruption of the Sun's ubiquitous magnetic field \-- the magnetism which shapes the heliosphere, moderates the energy to fill it in addition to providing the energy necessary for life on our planet. Probing that ocean and understanding the magnetism of the solar interior, its production, evolution, and eventual destruction is a huge challenge and is the primary scientific focus of our community.

The forced (magneto-)convection of the solar interior leaves visible tracers on the optical surface, the granular and super-granular scales that are a few and a few tens of megameters in diameter, respectively and the sunspots and active regions that can grow to several hundred megameters in size. However,  granulation and supergranulation are only the tip of the proverbial (magnetic) iceberg \citep[e.g.,][]{Nordlund2009}. A tertiary scale of convection has tantalized the community, that of ``giant cell convection.'' A rotationally-forced convective scale \citep[e.g.,][]{Wilson1987} which was first realized in the pioneering calculations of \citet{1968ZA.....69..435S} and \citep{Gilman1975} that may play a critical role in the formation of sunspots and active regions \citep[e.g.,][]{Weber2012}. Giant cells are hypothesized to reach the bottom of the convective interior and span one to two hundred megameters in diameter \citep[see the review of][for an extensive discussion]{Miesch2005,Nordlund2009}, but they remained an elusive quarry to the observer with only scarce hints of their existence \citep[e.g.,][]{1998Natur.394..653B}. A recent analysis which tracked the motion of supergranules in long Dopplergram timeseries indicated that a large\--scale pattern could be discerned from the data and that the pattern demonstrated behavior consistent with that expected of giant convective cells \citep[][]{Hathaway2013}.

It should be noted that contemporary investigations performed with high resolution observation and numerical simulations of solar surface convection strongly suggest that these apparently nested surface scales (with a strong emphasis towards the readily amenable granules and super-granules) reflect the multi-scale organization (or continuous spectrum) of motions present in the Sun's interior \citep[e.g.,][]{Berrilli2012, Orozco2012, Yelles2012, Yelles2014}. As such the third scale would indeed reflect a slower, longer-lived, convective scale present in the deepest interior \citep[e.g.,][]{Nordlund2009}.

In the following sections we present the application of an analysis technique that was originally designed to investigate the range of connective length scales of the outer solar atmosphere that could be inferred from a simple analysis of photospheric magnetograms \-- the ``Magnetic Range of Influence'' \citep[MRoI;][]{2006ApJ...644L..87M}. The MRoI analysis demonstrates the presence of a tertiary convective scale and that is an ever-present in the magnetogram record of the Helioseismic and Magnetic Imager \citep[HMI;][]{2012SoPh..275..207S} of the {\em Solar Dynamics Observatory} (\sdo). Furthermore, we see that the magnetic elements which comprise the vertices of this larger scale convective pattern are a potential anchor-site for an ever-present feature of the solar corona---extreme-ultraviolet BrightPoints \citep[BPs;][]{1973SoPh...32...81V}.

\begin{figure*}
\epsscale{1.15}
\plotone{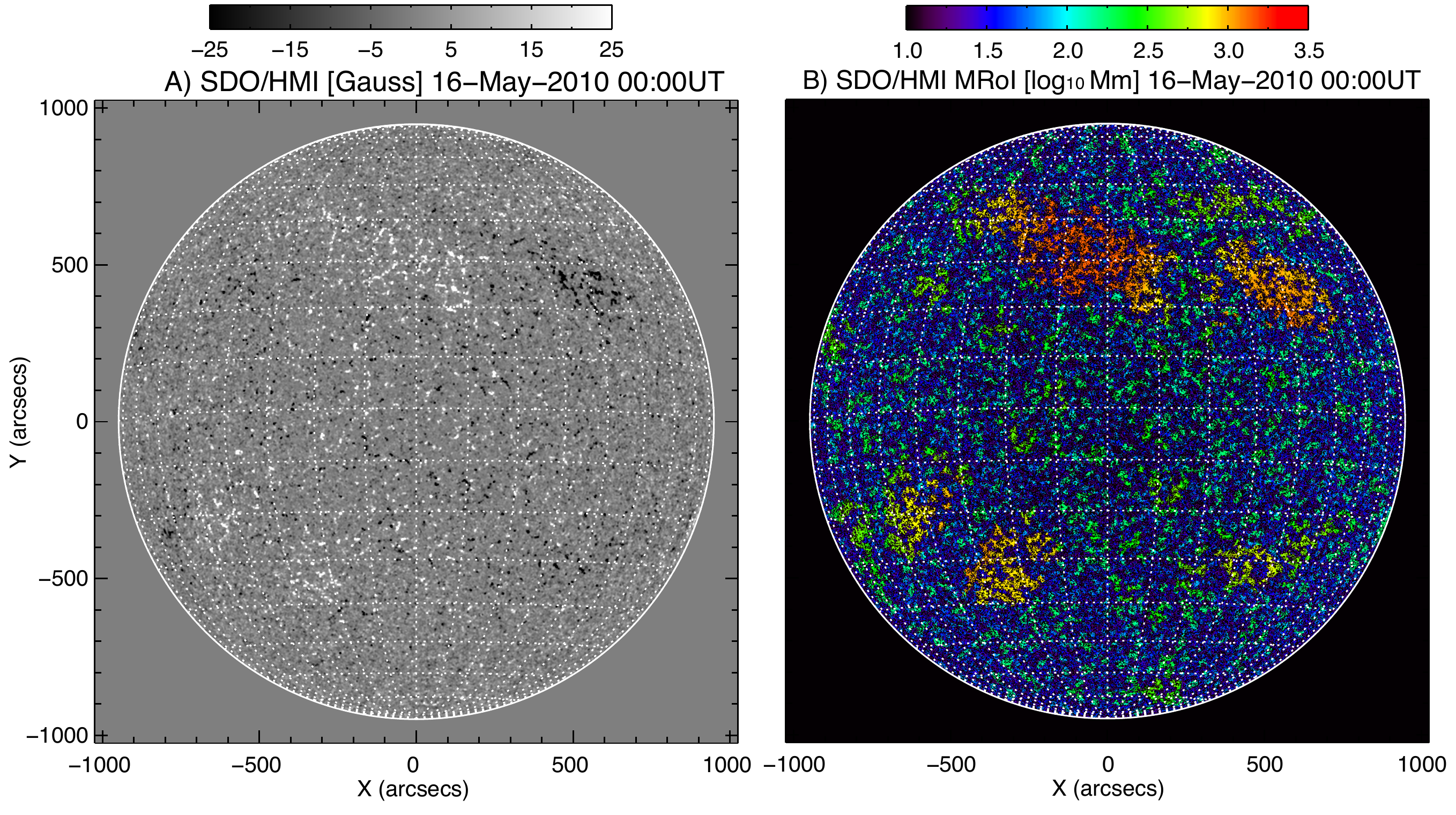}
\caption{The full-disk \sdo{}/HMI line-of-sight magnetogram (A) and derived ``Magnetic Range of Influence'' (MRoI; B) map from May 16 2010. \label{f1}}
\end{figure*}

\section{Analysis}\label{method}
\begin{figure*}
\epsscale{1.15}
\plotone{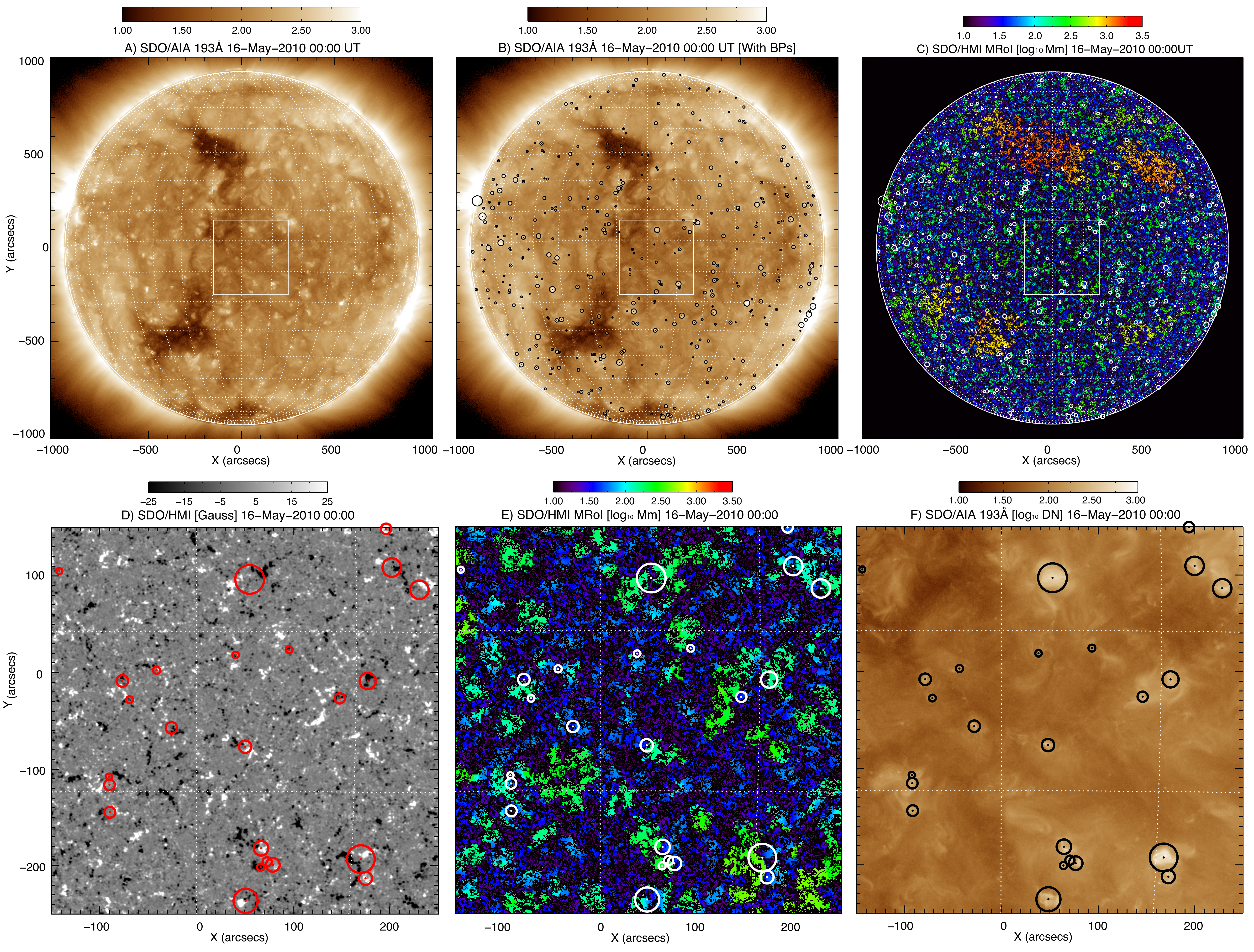}
\caption{Full disk \sdo{}/AIA image of the solar corona formed around 1.2MK in the 193\AA{} channel (A). Panel B shows the locations of the detected coronal BPs (black circles). Panel C shows the corresponding MRoI map (Fig.~\pref{f1}b) and the BP locations are white circles. The lower row of panels highlight the square inset region of the \sdo/HMI magnetogram, MRoI map, and coronal image with the brightpoints in the region identified as red, white and black circles respectively. \label{f4}}
\end{figure*}

A map of the magnetic range of influence, or MRoI, is constructed from a line-of-sight magnetogram in a pixel-by-pixel fashion and is defined as the (radial) distance the pixel over which the total signed flux of the enclosed region is zero. The MRoI is a measure of magnetic balance, or the effective length scale over which we would expect the overlying corona to be connected, or closed. The method was originally conceived to help identify the coronal holes \citep[][]{2006ApJ...644L..87M} which, as weak but spatially extended unipolar regions of the photosphere, should have large values of the MRoI. Conversely, in magnetically balanced regions like the quiet Sun MRoI values are small \citep[][]{McIntosh2007,2007ApJ...670.1401M}. Figure~\pref{f1} which shows the MRoI map (panel B) that is constructed from a ``720s'' \sdo/HMI line-of-sight magnetogram (panel A) taken at 00:00UT on May 16 2010 during a very quiet spell of solar activity. The most striking features in the MRoI map are the extended patches of MRoI values greater than 1,000 Mm in value \-- the orange-red regions, the spatially distributed patches of green and the blue-purple remainder (see the inset regions of Fig.~\pref{f4} for more detail). It should be noted that the present version of the MRoI does not account for spherical projection or line-of-sight projection of the line-of-sight field when computing the radial distance of closure. This leads to uncertainties in the MroI near the limb. Similarly, the fact that the visible disk can be dominated by large spatially-distributed unipolar patches \-- like coronal holes \-- leads to MRoI values which reach a solar diameter in scale. In this case there is no net-zero-field ``closure'' available at that largest scale.

\begin{figure}
\epsscale{1.15}
\plotone{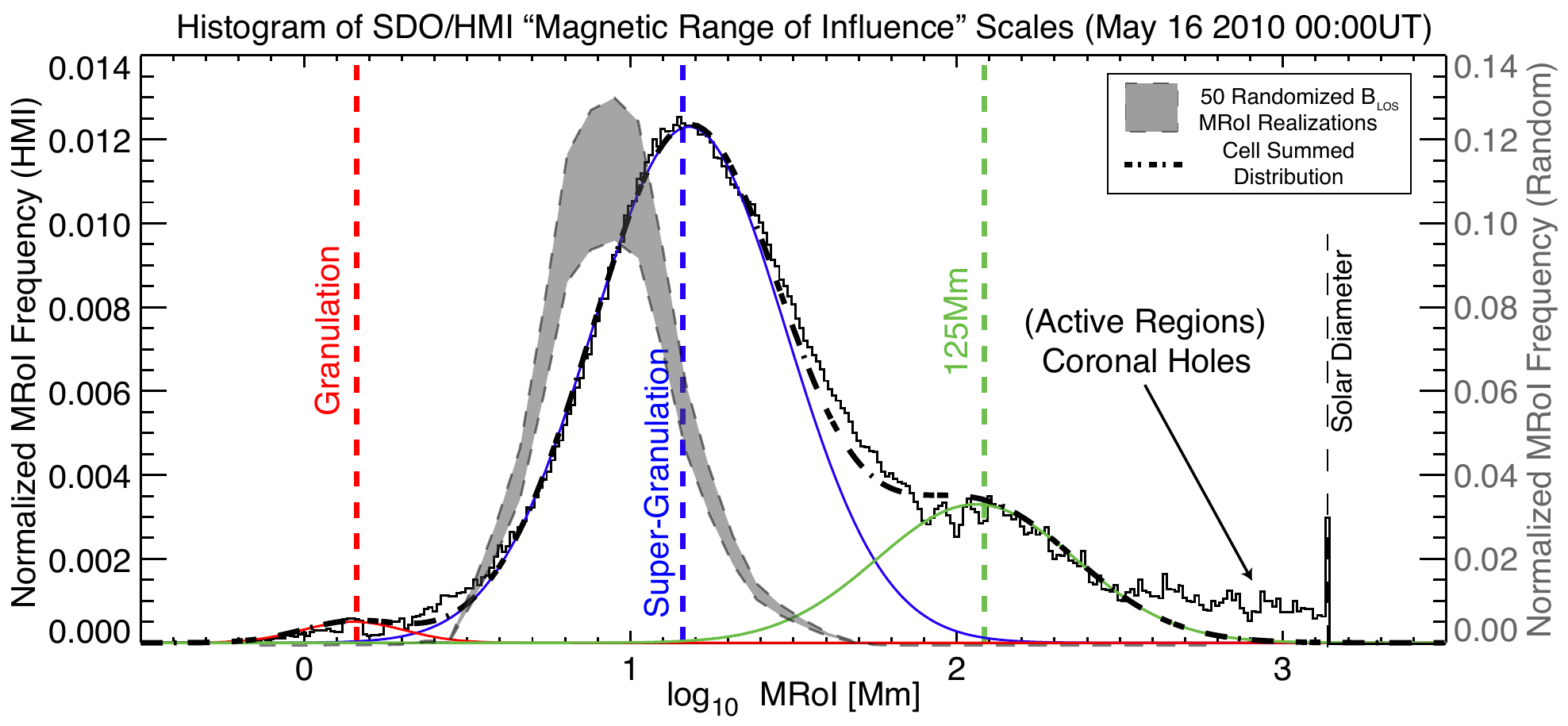}
\caption{Histogram of MRoI values derived from the full disk map of Fig.~\pref{f1} versus those of a ``null'' experiment (see text). The observed MRoI histogram shows four possible scale ranges: one near the resolution limit of \sdo{}/HMI at a few Mm (red dashed line), the majority of the pixels show a scale of $\sim$30Mm (blue dashed line), a third component peaking at $\sim$125Mm scale (green dashed line), and a group of scales around 1000Mm. The shaded ensemble of null MRoI histograms show none of the same characteristic scales. The MRoI has a hard limit of a solar diameter (vertical thin dashed line) imposed. \label{f2}}
\end{figure}

A histogram of the MRoI values from Fig.~\pref{f1} is shown in Fig.~\pref{f2}. There are four scales present in the histogram and (for illustrative purposes only) we have associated the three shortest scales with a Gaussian distribution and attempted to simultaneously identify a best-fit (dot\--dashed) line. The first component (red) is small in amplitude and, at just above the resolution limit of \sdo/HMI at a few Mm, is likely a faint signature of granulation. The second (blue) component of the MRoI distribution peaks around 30Mm and represents the magnetic scale of supergranulation. The third (green) component of the distribution is smaller in amplitude than that of the supergranules but seems to contain values in the 100 \-- 250Mm range and there is an excess distribution at very long scales which, as we will see below, have contributions from coronal holes and active region plage. While the first two and last component of the distribution are known, or identifiable, the third component of the distribution is composed of the spatially distributed green patches in Fig.~\pref{f1}B that have the 100 \-- 250Mm scale - a separation length scale that is commensurate with the expected dimension of giant convective cells \citep[e.g.,][]{1968ZA.....69..435S, 1998Natur.394..653B, Miesch2005}. For comparison we also show the distribution of scale histograms formed from the MRoI analysis of an ensemble of fifty randomized magnetograms (shaded region). Each null MRoI is computed using magnetogram values where the value in each pixel is drawn from an identical distribution to the original \sdo/HMI data. The shaded region in Fig.~\pref{f2} reflects the mean frequency and three standard deviations at each spatial scale. While this procedure demonstrates a profound peak of MRoI values around 8Mm we see {\em no} signature of smaller or larger scales in the randomized maps. This is a strong indication that the scales revealed by the MRoI analysis of the \sdo/HMI line-of-sight magnetograms reflect a range of characteristic connective length scales in the Sun's photosphere.

\begin{figure}
\epsscale{1.15}
\plotone{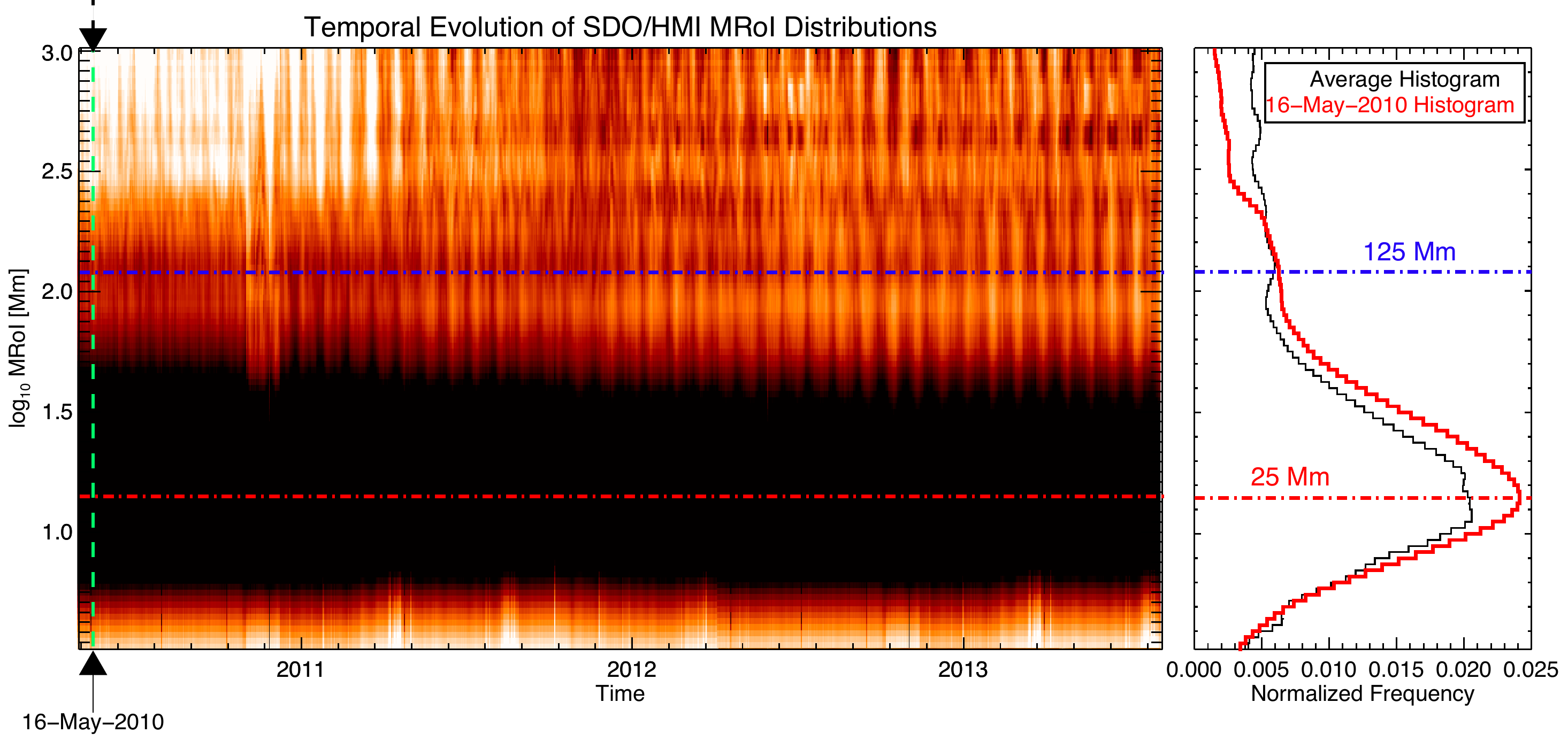}
\caption{The temporal evolution of the daily \sdo{}/HMI MRoI histograms from May 2010 to August 2013. Each vertical slice of the image is an MRoI scale distribution like that shown in Fig.~\pref{f2}. Representative supergranular and giant cell scales are drawn as red and blue horizontal dot-dashed lines respectively. The right panel shows the time averaged MRoI histogram (black) and the example from Fig.~\pref{f2} (red) from May 16 2010. \label{f3}}
\end{figure}

\subsection{Temporal Evolution of The Tertiary Scale}
Figure~\pref{f3} shows the variation of the MRoI length-scale distributions from May 2010 to August 2013. The inverse color image in the left panel shows shows that there are persistent features in the MRoI distributions, especially the components centered on $\sim$25Mm (the red horizontal dot-dashed line) and $\sim$125Mm (the black horizontal dot-dashed line). The right panel of the figure compares the temporal mean length-scale distribution (black line) with the May 16 2010 distribution of Fig.~\pref{f2} (red line). These measurements indicate that the $\sim$120Mm scale, commensurate with the expected scale of giant cells, is persistently visible in the solar photosphere using the MRoI technique. In addition to the prevalence of the $\sim$30Mm and $\sim$125Mm scales we see enhancements at $\sim$240Mm and $\sim$420Mm (or 2.375 and 2.625 log$_{10}$ MRoI respectively) which appear to grow in prevalence as cycle 24 proceeds. A simple assumption would be to associate these scales with the growth of sunspot complexes and coronal holes that are not present, or abundant, in the quiet 2010 spectrum. This interesting realization hints to a subtle solar cycle dependence in the scales which requires further investigation that is beyond the scope of this Letter, but will be explored in a subsequent publication (McIntosh et al.\ 2014 in preparation).

\begin{figure*}
\epsscale{1.15}
\plotone{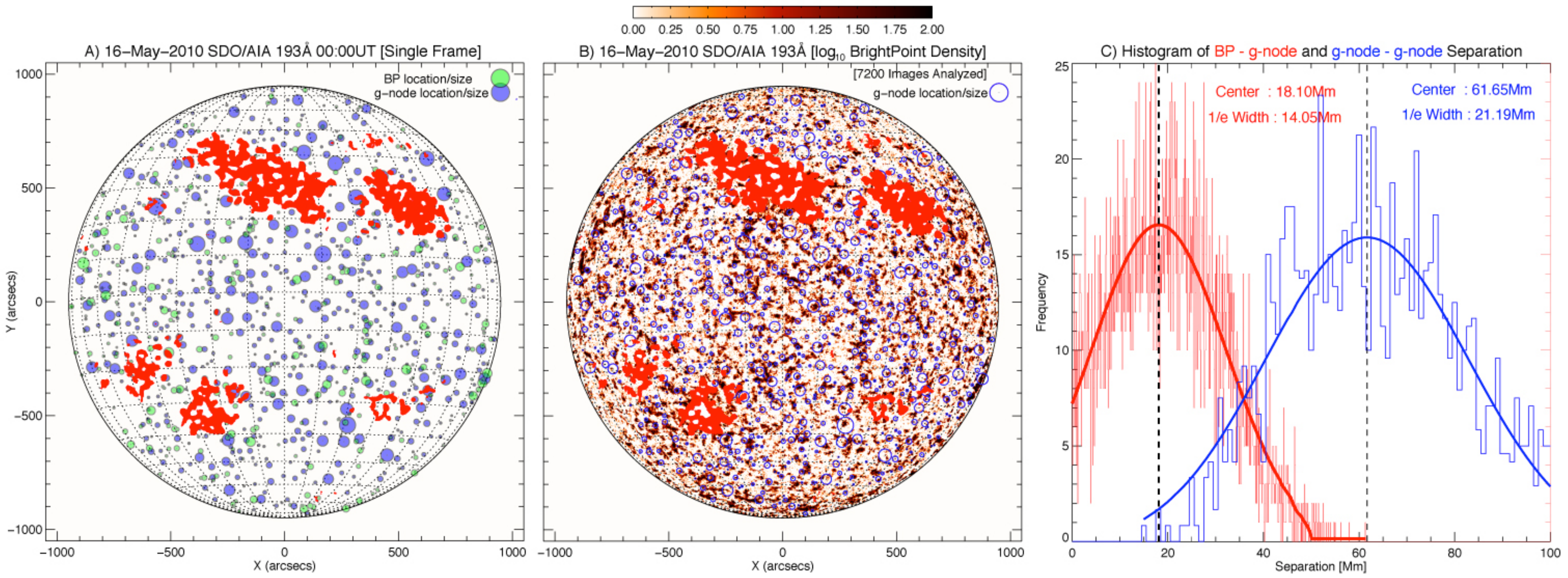}
\caption{From left to right the panels of the figure comparing the BP and g-node locations of May 16 2010. Panel A shows the sizes and locations of the identified g-nodes (blue filled circles) from the HMI magnetograms and the BPs (green filled circles) from the AIA 193\AA{} images. Panel B shows the BP location density composed from an analysis of 7200 AIA 193\AA{} images (sampled at 12s) in the 12 hours before and after 00:00UT on May 16 2010, the blue circles indicate the 00:00UT g-node locations as shown in panel A. Panel C shows the histogram of BP and g-node (edge-to-edge) separation for the five nearest BPs to each g-node (red distribution) and the same for g-node to g-node separation (blue distribution). \label{f5}}
\end{figure*}

\subsection{The Tertiary Scale and EUV Brightpoints}
Brightpoints, or BPs, are small ubiquitous constituents of the Sun's coronal plasma \citep[e.g.,][]{1973SoPh...32...81V}. It was originally noted that they are spatially associated with dipolar magnetic configurations \citep[e.g.,][]{1974ApJ...189L..93G} however, in a recent investigation \citet{2007ApJ...670.1401M} used the MRoI technique to demonstrate that BPs typically surround magnetic regions with a magnetic scale that is considerably larger than typical quiet sun values ($\sim$30Mm).

The upper panels of Fig.~\pref{f4} contrast coronal images and the MRoI. Panel A shows the \sdo{} Atmospheric Imaging Assembly \citep[AIA,][]{2012SoPh..275...17L} 193\AA{} image of the solar corona taken at 00:00UT on May 16 2010. Panel B shows the same image with the BPs identified using the method of \citet{2005SoPh..228..285M} as black circles, and panel C shows the corresponding MRoI map where the BPs are drawn as white circles.

Details of the BPs detection and tracking algorithms are available in the literature \citep[see, e.g.,][]{2005SoPh..228..285M, 2003ApJ...589.1062H}, although we have taken steps to improve the reliability of the detection in subsequent years that will be detailed in a forthcoming article which also presents the 1996 \-- 2014 BP record (McIntosh et~al.\ 2014 in preparation). Following image calibration and cosmic ray removal we construct a ``background'' image ($I_{b}$) using a 40Mm$\times$40Mm smooth version of that image ($I$). BPs are defined as spatially small (2-20Mm) three-$\sigma$ enhancements of the original image over the background image. That is, we construct a ``sigma'' image ($(I - I_{b}) / \sqrt{I_{b}}$) which can account for subtle differences of coronal images from instrument to instrument (e.g., say from \soho/EIT and \sdo/AIA) and provides significantly more robust BP determination than originally presented by \citet{2005SoPh..228..285M}. From the resulting histogram of sigma image values we use one, two, and three standard deviations above the mean value as the thresholds for BP detection \-- the three-$\sigma$ detections being the most reliable. After defining the detection thresholds we isolate contiguous pixel groups in the images, computing their center position and radius of gyration. Only three\--$\sigma$ regions with radii between 2 and 20Mm are considered as belonging to BPs and are those shown in Fig.~\pref{f4}.

On first inspection of Fig.~\pref{f4} we note that the BPs are well separated and that there appears to be a correspondence between regions of large MRoI and BP formation, as first noted by \citep[][]{2007ApJ...670.1401M}. The BPs appear to form in the vicinity of the green ($\sim$125Mm scale) concentrations that are spread over the solar disk. Exploring in a little more detail using the lower row of panels in Fig.~\pref{f4} we highlight the correspondence between BPs, the distribution of quiet sun magnetism, and the MRoI in more detail than visible in the full disk images above. First, we note from panel D that BPs do not form around all quiet Sun dipolar field pairs and that they do indeed appear to have preferred locations in the immediate neighborhood of the $\sim$125Mm-scale MRoI (green colored) regions. Hereafter, we will refer to these concentrations of $\sim$125Mm scale as ``g-nodes'' given their potential connection to the giant convective scale. We should note that a cursory inspection of the magnetogram would not naturally lead one to disentangle the magnetic flux at a supergranular vertex from that of a g-node \-- to all intents and purposes they are indistinguishable when using the magnetogram alone.

Figure~\pref{f5} demonstrates the apparent relationship between g-nodes and BPs (panel A). The g-nodes are identified from the MRoI maps using a modification of the BP detection algorithm---employing it to isolate only the 100-200Mm scale concentrations. In general, we see that BPs recurrently form in the immediate vicinity of g-nodes, but not all g-nodes have neighboring BPs (panel B). The general proximity of BP and g-node suggests that one ``pole'' of the BP's magnetic root is a g-node while the other is most likely a supergranular vertex of opposite magnetic polarity. This picture is supported by the average spatial separation of g-node and BP being $\sim$18Mm (panel C; red distribution) which is smaller than a typical a supergranular diameter, while g-nodes are considerably further apart ($\sim$62Mm; blue distribution). This preferred scale of BP formation supports observations that not all magnetic dipoles have associated BPs \citep[e.g., Fig.~\pref{f4}D and][]{1974ApJ...189L..93G}. Finally, we note that g-node detection in coronal holes are significantly impaired by spatially extended regions of large MRoI as shown by the red masked regions of  of the figure. Future versions of the MRoI analysis will address this long length-scale masking.

\section{Summary}\label{discuss}
Our observations have demonstrated that length scales of order 30Mm and 100 \-- 250Mm are readily discernible from MRoI diagnostics of photospheric line-of-sight magnetograms. This is possible because our MRoI technique infers the connective length scales of the outer atmospheric plasma from the photospheric magnetograms \-- a departure from the conventional (largely spectral) scale analysis \citep[e.g.,][]{Nordlund2009}. We anticipate that using a technique like MRoI to study numerical simulations of magneto\-convection (Rempel 2014, in preparation) may help to understand the correspondence between apparent connective scales demonstrated herein and the apparently scale-free environments inferred from other analyses of high resolution, small field-of-view simulations and observations \citep[e.g.,][]{Nordlund2009, Berrilli2012, Orozco2012, Yelles2012}. This pair of scales are visible throughout the record of \sdo/HMI line-of-sight magnetograms and the latter appears to be commensurate with the implied depth of the solar convection zone from calculations and helioseismic measurements \citep[e.g.,][]{1968ZA.....69..435S,Gilman1975,Wilson1987,1998Natur.394..653B}. Therefore, we infer that the 100 \-- 250Mm scale apparent in the MRoI maps identifies a pattern consistent in magnitude with that expected from giant convective cells. An extension of this work noted that EUV BPs appear to form preferentially around the magnetic vertices of this scale, features that we have dubbed g-nodes. The connection between these magnetic concentrations and EUV (and X-Ray) BPs has possibly been evident for some time. \citet{1978ApJ...219L..55G} noted that the longest lived BPs rotated at rates equivalent to much deeper rooted magnetic objects than surface phenomena. Further, it is possible that an archival survey of observations can be used to infer the presence of BPs and g-nodes that may trace back to the pioneering work of Karen Harvey and others linking ephemeral active regions, BPs, the torsional oscillation, and other phenomena connected to the solar cycle \citep[e.g.,][]{1988Natur.333..748W, 1992ASPC...27..335H}.

The methods highlighted above offer the potential of a near-real-time mapping of the Sun's largest convective scale. This scale, driven by the rotation of the convective plasma ocean \citep[][]{Gilman1975,Wilson1987,Miesch2005}, was also recently uncovered using Dopplergrams from \sdo/HMI during the same very quiet epoch of solar activity \citep[][]{Hathaway2013}. Understanding, and possibly combining, these new analysis techniques \citep[also including the recently discovered, and potentially related, ``coronal cells'';][]{2012ApJ...749...40S} could yield a significant breakthrough in understanding of the solar interior and solar cycle. An investigation combining the \soho{} and \sdo{} epoch observations is forthcoming (McIntosh et al.\ 2014 in preparation).

\section{Conclusion}\label{conclude}

We have seen that the quiescent photospheric magnetic field is composed of multiple connective scales. The observed scales range from a few megameters to those that are 100 \-- 250Mm in scale. We expect that the latter of these scales belongs to a spatially large, deep and hence slowly overturning convective flow \-- one that possibly reaches to the bottom of convection zone. Further, it would appear that photospheric line-of-sight magnetograms (and Dopplergrams) carry information about these nested scales in a non-trivial spatial mixture. It follows that the two cannot be easily disentangled without employing a technique like the MRoI. However, the ready visibility of a giant convective scale and its relatively straightforward identification could have a significant bearing on our ability to probe the variations of the deep solar interior and its long-term evolution.

\acknowledgements
We thank Guiliana de Toma and the anonymous referee for their helpful comments on the manuscript. NCAR is sponsored by the National Science Foundation. We acknowledge support from NASA contracts NNX08BA99G, NNX11AN98G, NNM12AB40P, NNG09FA40C ({\em IRIS}), and NNM07AA01C ({\em Hinode}).

\end{document}